\documentclass[pt10,journal=jacsat,manuscript=article]{achemso}
\pdfoutput=1
\setkeys{acs}{articletitle = true}

\usepackage{units}
\usepackage{graphicx}
\usepackage{amsmath}
\usepackage{esint}
\usepackage{amssymb}   
\usepackage{breqn}
\usepackage{achemso}
\usepackage[version=3]{mhchem}
\usepackage{titlesec}
\usepackage{enumitem,xcolor}
\usepackage{bm}
\usepackage{mathtools}
\usepackage{multirow}

\definecolor{stefano_green}{rgb}{0.31,0.53,0.10}
\definecolor{stefano_red}{rgb}{0.85,0.23,0.11}
\definecolor{stefano_blue}{rgb}{0.0,0.0,0.99}
\definecolor{NM}{rgb}{0.5,0.,0.5}

\titleformat{\section}[display]{\bfseries}{}{0.0ex}{}[] 
\titleformat{\subsection}[runin]{\bfseries}{}{0.0ex}{}[] 
\titleformat{\subsubsection}[runin]{\itshape}{}{0.0ex}{}[]

\makeatletter

\let\oldmaketitle\maketitle
\let\maketitle\relax

\newcommand{\onlinecite}[1]{\hspace{-1 ex} \nocite{#1}\citenum{#1}} 

\newcommand\mvec[1]{\mathbf{#1}}  

\def\IFC{{\small IFC}}  \def\IFCs{{\small IFC}s}  
\def\DFT{{\small DFT}} 
\def\ML{{\small ML}} 
\def\RF{{\small RF}}  \def\RFs{{\small RF}s}  
\def\VASP{{\small VASP}} 
\def\USPEX{{\small USPEX}} 
\def\PAW{{\small PAW}} 
\def\sv{{\substack{\scalebox{0.6}{V}}}} 
\def\svib{{\substack{\scalebox{0.6}{vib}}}} 
\def\smax{{\substack{\scalebox{0.6}{max}}}} 
\def\sat{{\substack{\scalebox{0.6}{atoms}}}} 
\def\sopt{{\substack{\scalebox{0.6}{opt}}}} 

\title{Vibrational properties of metastable polymorph structures by machine learning} 

\author{Fleur Legrain}
\email{fleur.legrain@cea.fr}
\affiliation[CEA, LITEN]
{CEA, LITEN, 17 Rue des Martyrs, 38054 Grenoble, France}
 
\author{Ambroise van Roekeghem}
\affiliation[CEA, LITEN]
{CEA, LITEN, 17 Rue des Martyrs, 38054 Grenoble, France}
\author{Stefano Curtarolo}
\affiliation[{Duke University}]
{Department of Mechanical Engineering and Materials Science, Duke University, Durham, North Carolina 27708, United States;
Fritz-Haber-Institut der Max-Planck-Gesellschaft, 14195 Berlin-Dahlem, Germany}
\author{\\Jes\'{u}s Carrete}
\affiliation[TUWien]
{Institute of Materials Chemistry, TU Wien, A-1060 Vienna, Austria}
\author{Georg K. H. Madsen}
\affiliation[TUWien]
{Institute of Materials Chemistry, TU Wien, A-1060 Vienna, Austria}
\author{Natalio Mingo}
\email{natalio.mingo@cea.fr}
\affiliation[CEA, LITEN]
{CEA, LITEN, 17 Rue des Martyrs, 38054 Grenoble, France}
\makeatother

\usepackage[english]{babel}
\usepackage{color}
\begin{document}
\oldmaketitle
\begin{abstract}
  \noindent
    Despite vibrational properties being critical for the {\it ab initio} prediction of the finite temperature stability and transport properties of solids,
    their inclusion in  {\it ab initio} materials repositories has been hindered by expensive computational requirements.
    Here we tackle the challenge, by showing that a good estimation of force constants and vibrational properties 
    can be quickly achieved from the knowledge of atomic equilibrium positions using machine learning.
    A random-forest algorithm trained on only 121 metastable structures of KZnF$_3$ reaches a maximum absolute error of 0.17 eV/$\textrm{\AA}^2$ for the  interatomic force constants, 
    and it is much less expensive than training the complete force field for such compound.
    The predicted force constants are then used to estimate phonon spectral features, heat capacities, vibrational entropies, and vibrational free energies, 
    which compare well with the {\it ab initio} ones.
    The approach can be used for the rapid estimation of stability at finite temperatures.
\end{abstract}

\section{Introduction}
  Large databases of calculated material properties, such as {\sf AFLOW}.org\cite{aflowPAPER,aflowlibPAPER}, 
  the Materials Project\cite{Jain2013}, and
  {\small OQMD}\cite{OQMD},  
  have become powerful tools for accelerated materials design\cite{nosengo_can_2016,Thehigh-throughputhighwaytocomputationalmaterialsdesign,doi:10.1063/1.4977487}. 
  {\it Ab initio} relaxed crystal structures and ground state energies are routinely provided in these repositories, and often used to evaluate phase diagrams starting from zero temperature
  or with simple approximations \cite{curtarolo:art115}. With this approach roughly 50\% of the experimentally known compounds are found above the convex hull.\cite{Opahle_NJP13,sun_thermodynamic_2016} This can be due to the experimental structure being truly metastable. Another possible explanation could be the lack of accuracy of standard density functional approximations.\cite{Sun_NatChem16} However, an important factor will undoubtedly be that phonon-related contributions are highly important at the temperatures of interest\cite{liu_first-principles_2017,burton_first-principles_2006,r._akbarzadeh_first-principles_2007,carrete_structural_2017,phononbroadeninginHEA}. These contributions are often neglected, principally due to the high computational cost posed by the interatomic force constants (\IFC) matrix, {\it i.e.}the Hessian, or second derivatives of the energy with respect to the atomic displacements. Similarly, structural global energy minimization methods, such as {\small USPEX}\cite{doi:10.1063/1.2210932,GLASS2006713}, generate hundreds of relaxed candidate structures. However, both for large databases and global energy methods, the vibrational energy contributions are typically too expensive to be calculated with brute force. 
  A considerable advantage would 
  come from
  an on-the-fly estimation of vibrational free energies during the search.
  
Neglecting phonon contributions to the free energy is obviously wrong and this practice is mainly due to computational necessities. Obtaining the Hessian typically requires one or two orders of magnitude more computer time than the corresponding structural relaxation. However, neglecting phonons can have dramatic consequences. For example, 
  vibrational contributions have been shown to modify the sequence of reactions occurring as a function of temperature or pressure\cite{r._akbarzadeh_first-principles_2007}, 
  to explain the precipitation sequence of metallurgical phases\cite{wolverton:prl_2001_AlCu}, 
  or to alter the stability ordering of novel 2D material phases\cite{carrete_structural_2017}. 
  Phonons also have been shown to be as important as configurational disorder for the prediction of alloy phase diagrams and thereby essential to obtain experimental agreement\cite{liu_first-principles_2017,burton_first-principles_2006}. 
  Particularly relevant is the problem of polymorphs, i.e. materials sharing the same chemical composition but having different crystal structures.
  Calculations on organic molecules have shown that $\sim$69\% of polymorph pairs reversed their relative stability when increasing the temperature, due to the vibrational contribution to the free energy\cite{C5CE00045A}. 
Also, roughly 50\% of the compounds in the Materials Project database are metastable with a median energy above the convex hull of 15 meV/atom\cite{sun_thermodynamic_2016}
  and similar values apply to the {\small ICSD} \cite{ICSD} repository within {\sf AFLOW}.org.
  This energy is comparable to typical phonon free energy differences between polymorphs\cite{van_de_walle_effect_2002,wolverton:prl_2001_AlCu,curtarolo_accuracy_2005}, 
  highlighting the importance of including the phonon vibrational energy when determining the finite temperature ground states.
  The high-throughput prediction of phase diagrams at finite temperatures is still a major challenge for computational materials design, 
  mostly because of the difficulty to quickly compute Hessians \cite{Thehigh-throughputhighwaytocomputationalmaterialsdesign,doi:10.1063/1.4977487}. 
  Clearly, there is an urgent need for a rapid and reliable approach to predict the \IFCs.

  Machine learning (\ML) algorithms can be used to avoid costly calculations.
  \ML\ has been successfully used to predict \IFCs\ for compounds from the same crystal structure but different chemical composition \cite{PhysRevX.4.011019,PhysRevX.6.041061},
  which was subsequently shown to be a major factor determining the vibrational free energy of compounds\cite{doi:10.1021/acs.chemmater.7b00789}. 
However, the more complex problem of predicting \IFCs\ of competing structures of the same composition has not been addressed. 
  Often the relaxed structures are already known and the challenge is to predict only the computationally expensive Hessians. 
  This is the case with large {\it ab initio} databases, which contain many metastable structures or artificial configurations for sampling the phase space \cite{aflowlibPAPER}.
  Contrary to force-field fitting where a continuum of ``deformations$\rightarrow$forces'' states has to be sampled,
  here accurate representations of the potential energy surface around diverse and potentially uncorrelated metastable states is needed. Is this doable?
  In the present work we tackle the challenge by finding
  an efficient solution with the help of random forests, trained with only one hundred metastable structures, but still capable of predicting accurate \IFCs, spectral properties and thermodynamic quantities.

\section{Approach}
The interatomic force constants between atom $i$ and $j$ constitute a second-order tensor defined by the second derivatives of the PES with respect to atomic displacements
\begin{eqnarray}
    \mathbf{\Phi}_{ij}=(\nabla_{\!\mvec{r}_i}\otimes\nabla_{\!\mvec{r}_j}) E
  \end{eqnarray}
For \ML\ to predict the $\mathbf{\Phi}_{ij}$'s we need to construct atom-centered descriptors based on an internal coordinate representation that is invariant with respect to the symmetries of the systems, as well as permutations among atoms of the same species. A similar challenge is faced in force-field fitting\cite{doi:10.1063/1.3553717,PhysRevB.92.094306,QUA:QUA24836} but here we face the additional problem of generalizing the concept to tensors.

Scalar quantities of the physical system, like the energy, are expressed in this representation as functions of a set of scalar descriptors, $\{g_{i,j}^{\alpha}\}$, based on these internal coordinates. 
  Vector quantities associated to the $i^{\textrm{th}}$-atom can similarly be expressed by descriptors 
$ g_{ij}^\alpha\mvec{r}_{ij}$, that transform contravariantly.
  More generally, however, one can produce quantities that transform as tensors by taking gradients of the scalar descriptors.

 We choose a series of Gaussians, similar to those used in force-field fitting\cite{doi:10.1063/1.3553717}, to represent the pair part
  \begin{eqnarray}
    g_{ij}^\alpha=e^{-\left({\mvec{r}_{ij}\over a_\alpha}\right)^2}
  \end{eqnarray}
  where $\left\{a_\alpha\right\}$ are a set of radii spanning a few interatomic distances encompassing atoms $i$ and $j$. Taking the gradients of these scalar descriptors leads to $3\times3$ matrices defined for each atomic pair $(i,j)$ as:
  \begin{eqnarray}
    M^{\eta,\eta'}\equiv {\partial^2 \over \partial {x_i}^\eta \partial {x_j}^{\eta'} }g_{ij}^\alpha
    = {2\over a_\alpha^2} g_{ij}^\alpha \left[ \delta^{\eta,\eta'}-{{2r^\eta_{ij} r^{\eta'}_{ji}}\over{a_\alpha^2}}\right],
  \end{eqnarray} 
  where $\eta$ and $\eta'$ run over the three Cartesian coordinates.
  While the $\delta$ term transforms as a scalar, the $r_{ij}^\eta r_{ji}^{\eta'}$ term corresponds to the outer product of the gradients of scalar function $g$ and transforms as a rank-2 tensor. 
  Therefore, descriptors of the type 
  \begin{eqnarray}
    \mathbf{D}_{ij}^{(2)\alpha}=(\nabla_{\!\mvec{r}_i}\otimes\nabla_{\!\mvec{r}_j})g_{ij}^\alpha     \propto g_{ij}^\alpha\mvec{r}_{ij}\otimes \mvec{r}_{ji},
  \end{eqnarray} 
  transform as rank-2 tensors and can be used for the regression of Hessians.
   Periodic boundary conditions within the supercell spanning the force cut-off range (here  $5 \times 5 \times 5$) require an extra modification of the descriptor as
    \begin{equation}
      {\mathbf D}_{i,j}^{(2)\alpha} = \sum_{m}e^{-\left|{\mvec{r}_{ij}+\mvec{R}_m\over a_\alpha}\right|^2}(\mvec{r}_{ij} + \mvec{R}_m) \otimes (\mvec{r}_{ji} - \mvec{R}_m),
      \label{periodicdescriptor}
    \end{equation} 
where $\mvec {R}_m$ are the translation vectors connecting identical atoms in the supercell.

The set of descriptors above can be extended to higher orders, at an increased computational expense. For instance, the following set of rank-2 tensor descriptors would capture further 3-body interactions.
  \begin{eqnarray}
    {\mathbf D}_{ij}^{(3)\alpha,\beta,\gamma} &=&\sum_k \{ g_{ik}^\alpha g_{kj}^\beta\nabla_{\!\mvec{r}_i}\theta_{ikj}\otimes \nabla_{\!\mvec{r}_j}\theta_{ikj}
                                                  + g_{ij}^\gamma g_{jk}^\beta\nabla_{\!\mvec{r}_i}\theta_{ijk}\otimes \nabla_{\!\mvec{r}_j}\theta_{ijk} + g_{ki}^\alpha g_{ij}^\gamma\nabla_{\!\mvec{r}_i}\theta_{kij}\otimes \nabla_{\!\mvec{r}_j}\theta_{kij}\} \nonumber \\
                                              &\equiv& \sum_k {\mathbf D}^{\alpha,\beta,\gamma}_{i,k,j}.
  \label{phi3}
  \end{eqnarray}
  where $\theta_{ijk}$ is the angle formed by atoms $i$,$j$ and $k$.
  The gradients of $\theta_{ijk}$ can be expressed in terms of cross products of pairs in $\{i,j,k\}$ and ${\mathbf D}_{ij}^{(3)}$ transform as a tensor.
    There are other ways to define descriptors involving two and three-atom terms\cite{doi:10.1063/1.3553717,artrith_efficient_2017}. 
  However, to the best of our knowledge, direct regression of Hessians using invariant tensorial-form descriptors has not been attempted before.

  Descriptors $D_{i,j}^{(2) \alpha}$ are used to predict $\mathbf \Phi_{ij}$, i.e. the $3 \times 3$ matrices of \IFCs\ between different $i$- and $j$-atoms.
  Atomic force constants of the form, $\mathbf \Phi_{ii}$, which simply describes the forces on atom $i$ due to its own displacement, cannot come from $D_{i,i}^{(2) \alpha}$.
  Thus, the whole environment is included through the sum:
  \begin{dmath}
    {\mathbf D}_{i,i}^{({\mathrm{diag}})(2)\alpha} = \sum_{j} {\mathbf D}_{i,j}^{(2)\alpha},
     \label{diagonaldescriptor}
 \end{dmath}
  \noindent
  where $j$ indices through all the atoms of the {\bf supercell} , including $i$ itself.
  To get the whole \IFCs, two different \ML\ models are then trained:  $D_{i,j}^{(2) \alpha} \xRightarrow{\text{\footnotesize ML}} \mathbf \Phi_{i\neq j}$ and $D_{i,i}^{({\mathrm{diag}})(2)\alpha} \xRightarrow{\text{\footnotesize ML}} \mathbf \Phi_{ii}$.

  For clarity of the presentation, the dependence on chemical species, required for multi-component systems, has not been included in previous the formulas.
  Given a set of species $\{s\}$, the descriptors can be written as 
  ${\mathbf D}_{s,s';i,j}^{(2) \alpha}\equiv (\delta_{s_i,s}\delta_{s_j,s'}+\delta_{s_i,s'}\delta_{s_j,s}){\mathbf D}_{i,j}^{(2)\alpha}$, and ${\mathbf D}^{(3)\alpha,\beta,\gamma}_{s,s',s'',i,j}=\sum_k \delta_{s_k,s''}(\delta_{s_i,s}\delta_{s_j,s'}+\delta_{s_i,s'}\delta_{s_j,s}) {\mathbf D}^{(3)\alpha,\beta,\gamma}_{i,j,k}$,
  with $s$ and $s'$ species indices.

\section{Results and discussion}

\subsection{Data set.} 
  The \ML\ approach is developed for a test chemical system: the metastable structures of KZnF$_3$ (a cubic perovskite at 0K, chosen for simplicity).
  The initial data set consists of 267 KZnF$_3$ structures with 10 atoms per unit cell, randomly generated by the first-generation run of the \USPEX\ code\cite{doi:10.1063/1.2210932,GLASS2006713}, 
  and optimized using density functional theory (\DFT)\cite{InhomogeneousElectronGas,PhysRev.140.A1133} as implemented in \VASP\ \cite{Efficiencyofab-initiototalenergy}.
  The projector augmented wave (\PAW) method is employed to deal with the core and valence electrons\cite{Fromultrasoftpseudopotentials}. 
  The data sets preparation follows {\sf AFLOW.org} high-throughput recommendations \cite{aflowPAPER,RESTful_API_2014,curtarolo:art104}, 
  and the kinetic energy cutoffs are set to 450 eV for the plane wave basis.
  The force constant matrices and the phonon frequencies are computed at $\Gamma$ using density functional perturbation theory\cite{DFTP}.

  The identification and reduction of symmetrically equivalent cells is performed through the following structural fingerprint.
  For every structure $C$, descriptors are computed for each $i$-, $j$-atom pair:
  $D_{s,s';i,j}(C)^{\alpha} \equiv (\delta_{s_i,s}\delta_{s_j,s'}+\delta_{s_i,s'}\delta_{s_j,s}) g_{ij}^\alpha(C) $, where $\{s,s'\}$ are  the species indices
  (these same descriptors are also employed to predict the \IFC\ matrix invariants, as detailed later).
  The sum over the pairs leades to a fingerprint $K$ for a given structure $C$:  
  \begin{equation}
    K^\alpha_{s,s'}(C) \equiv \sum_{i,j} D_{s,s';i,j}^{\alpha}(C) = \sum_{i,j} (\delta_{s_i,s}\delta_{s_j,s'}+\delta_{s_i,s'}\delta_{s_j,s}) g_{ij}^\alpha(C)
  \end{equation}
  The distance $d(C_1,C_2)$ between structures $C_1$ and $C_2$ is then defined as: 
  \begin{equation}
    d(C_1,C_2) \equiv 
    \sum_{(s,s';\alpha)}
    \frac{| K^\alpha_{s,s'}(C_1)- K^\alpha_{s,s'}(C_2)|}{| K^\alpha_{s,s'}(C_1)|+| K^\alpha_{s,s'}(C_2)|}.
  \end{equation}
  \noindent
  After combinatorial analysis between the structures, an optimum distance's threshold of 0.35 is found by inspection.
  Only 121 inequivalent configurations are found, the remaining ones being discarded as duplicates (Supplementary Materials).
  The 121 cells are then used to build the \ML\ model and assess its performance.

\subsection{Predicting force constants.}
\begin{figure}
  
  \includegraphics[width=0.99\linewidth]{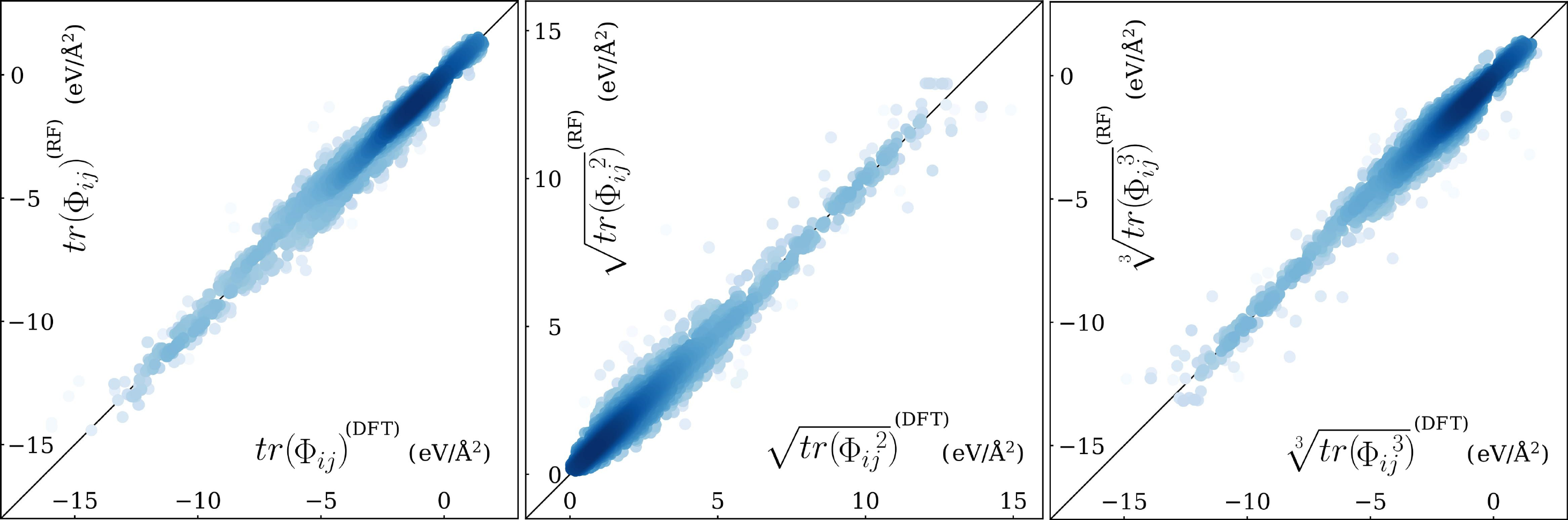}
  \caption{\small \small  
    Random forest performance: predictions {\it versus} calculations of force constant sub-matrix components across $i$- and $j$-atoms. 
    The abscissa shows the values obtained with \DFT, and the ordinate those obtained with \RFs.}
  \label{figure2}
 \end{figure}

  Amongst the available regression algorithms, random forests (\RF) are chosen because they are non-parametric, require virtually no data pre-conditioning, and usually yield robust and reliable results.
  The {\sf scikit-learn} implementation\cite{Pedregosa_JMLR_2011} is used to assess the performance of the model via 10-fold cross validation. 
  The forest contain 100 trees: better performance was not noticed with larger forests. 
  The three independent scalar invariants \cite{Longman}
  $\{ tr(\mathbf \Phi_{ij}), \sqrt{tr(\mathbf \Phi_{ij}^2)}, \sqrt[3]{tr(\mathbf \Phi_{ij}^3)} \}$
  --- derived from calculated or predicted Hessians and defined for each \IFC\ 
  between different atoms --- are used to assess the quality of the model.
  The performance of the random forests is listed in Table \ref{table1} and depicted in Figure~\ref{figure2}.

 \begin{table*}
  \caption{\small \small Performance: statistical analysis of \DFT\ calculations {\it versus} \RFs\ predictions.}      
  \label{table3}
  \footnotesize
  \begin{tabular}{||c||c|c|c|c||}
    \hline
    &  Pearson & Spearman & mean absolute& root mean \\
     &  coefficient & coefficient & error & square error \\ \hline
    {$tr(\mathbf \Phi_{ij})$}       &  0.99  & 0.98 & 0.25& 0.38 \\ 
            &    &   & (eV/$\textrm{\AA}^2$) & (eV/$\textrm{\AA}^2$) \\ \hline
    {$\sqrt{tr(\mathbf \Phi_{ij}^2)}$}       &   0.98     & 0.95 & 0.27& 0.41\\ 
            &    &   & (eV/$\textrm{\AA}^2$) & (eV/$\textrm{\AA}^2$) \\ \hline
    {$\sqrt[3]{tr(\mathbf \Phi_{ij}^3)}$}  &   0.98  & 0.95 & 0.28& 0.43\\ 
            &    &   & (eV/$\textrm{\AA}^2$) & (eV/$\textrm{\AA}^2$) \\ \hline
    $\mathbf \Phi_{ij}$ &  0.99   & 0.93 & 0.17& 0.32\\ 
            &    &   & (eV/$\textrm{\AA}^2$)  & (eV/$\textrm{\AA}^2$) \\ \hline
    variance &  0.88  & 0.87 & 0.88 & 1.14\\ 
    $\sqrt{\overline{\left(\omega-\overline{\omega} \right)^2}}$ &    & & (rad/ps) & (rad/ps) \\ \hline
    mean &   0.92   & 0.88 & 0.73 & 0.94\\ 
    $\overline{\omega}$ &    & & (rad/ps) & (rad/ps) \\ \hline
    max &  0.88  & 0.87 & 3.79 & 4.63\\ 
     $\omega_\smax$ &    & & (rad/ps) & (rad/ps) \\ \hline
    $C_v$ & 0.91    & 0.83 & 0.0008 & 0.0010\\ 
            &    &   & (meV/K/atom) & (meV/K/atom) \\ \hline
    $F_\svib$ &  0.82 & 0.80 & 2.92 & 3.78\\ 
            &    &   & (meV/atom) & (meV/atom) \\ \hline
    $S_\svib$ &  0.80 & 0.79 & 0.009 & 0.012 \\    
            &    &   & (meV/K/atom) & (meV/K/atom) \\ \hline
  \end{tabular} 
  \label{table1}
\end{table*}

  Upon trying with various different choices of radii $a_\alpha$, the best results are achieved with $a_\alpha=\{1,2,3,...,30\}$ $\textrm{\AA}$. 
The outcome of the descriptor with periodicity (Eq.~\eqref{periodicdescriptor}) is satisfactory.
  On the contrary, the performance is poor when periodicity is neglected (mean absolute errors are larger than 1eV/$\textrm{\AA}^2$, see Supplementary Materials).
  Errors are larger on small supercells, and decrease if training is performed on larger systems.

\begin{figure}

\includegraphics[width=0.6\linewidth]{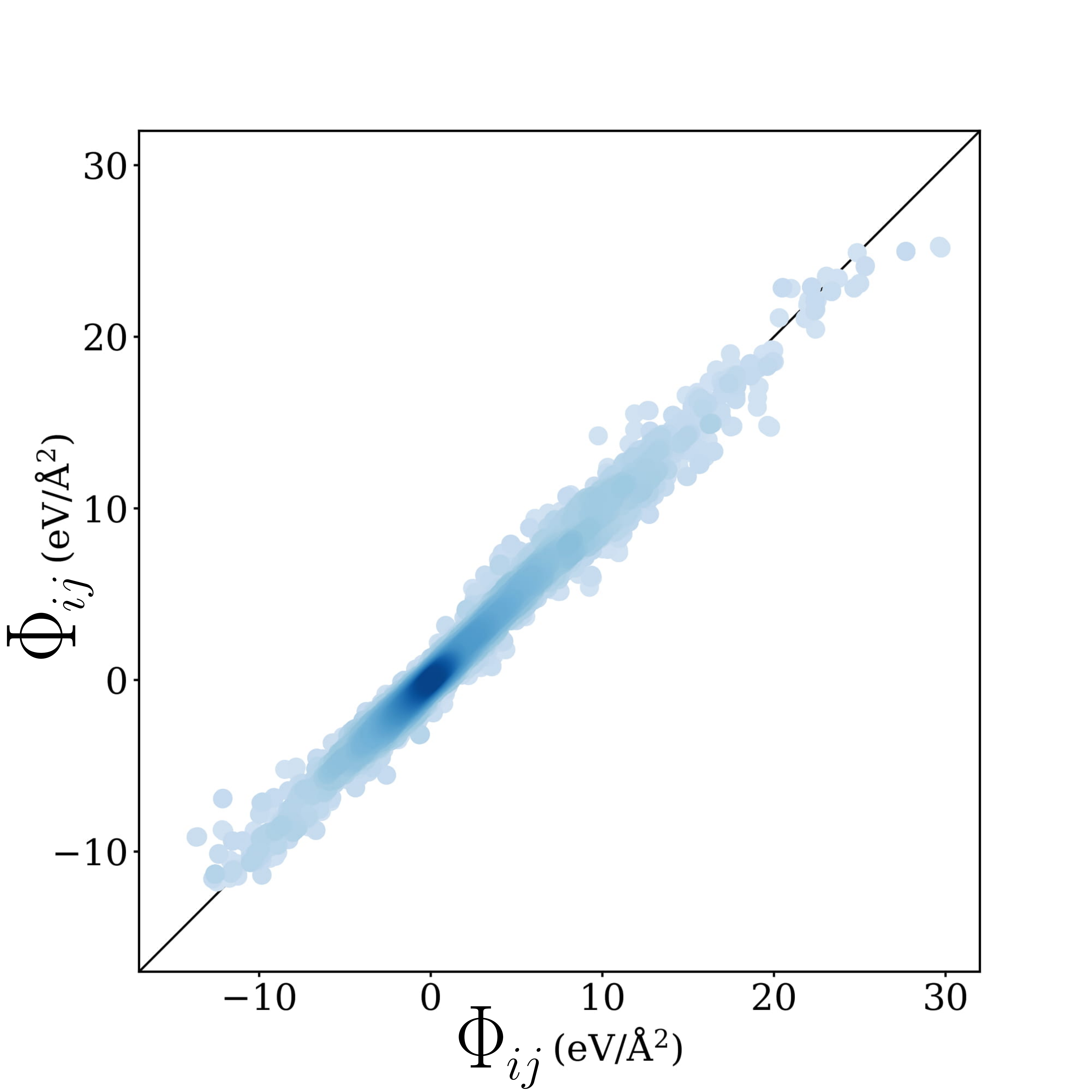}
\caption{\small \small  
  Random forest performance: predictions {\it versus} calculations for individual interatomic force constants. 
  The abscissa shows the values computed with \DFT\ and the ordinate shows the values obtained with \RFs.}
  \label{figure3}
\end{figure}

  The full Hessian is tackled with the descriptors from Eq.~\eqref{periodicdescriptor} 
and Eq.~\eqref{diagonaldescriptor}. 
The individual force constants predicted {\it versus} calculated --- are compared in Figure~\ref{figure3} and listed in Table~\ref{table1}.

  The \RF\ results indicate that very good predictions can be obtained using only simple two-atom descriptors.
  The extension to the 3-atom environments,  Eq.~\ref{phi3},  
  does not noticeably improve the outcome,
  implying that the key factors determining the \IFCs\ are the species and relative positions between atoms' pairs, without much contribution from other environmental atoms. 
  3-atom descriptors take much longer to calculate, are much more numerous than the pair descriptors, and therefore impose constraints onto the number of accessible radii $\{\alpha\}$, potentially leading to sub optimal results.
  Thus, it is possible that other descriptor algebraic formalisms and/or broader training sets -- more systems and larger structures --- could improve the outcome when
  3-atom environments are accessed. This is beyond the scope of the current work and it will be tackled in the future.

  How does the promising accuracy of predicted \IFCs\ translate into
  vibrational properties? 
  Errors do accumulate and even an apparently good
  prediction of forces could still violate conservation rules of the system leading to unphysical results.
  The phonon frequencies of the different KZnF$_3$ cells are computed
  from the \RF-predicted \IFCs. 
  Results are extremely sensitive to small inaccuracies in
  predicted forces and imaginary phonon frequencies appear.

  The  imaginary modes are removed by correcting each Hessian $H$ to the ``closest'' semipositive definite matrix $H'$.
  A diagonal matrix $D$ is obtained through the basis transformation $H=PDP^T$. A corrected $D \rightarrow D_{\rm c}$ is produced by 
  replacing the negative terms with zeroes. The object is rotated back to the original basis, $PD_{\rm c}P^T$.
  The term is further corrected by enforcing the acoustic sum rule
  leading to $H_{\rm c}=(I-Q) (PD_{\rm c}P^T) (I-Q)$, with $Q \equiv \sum_{t=1,2,3}v_t^{T} v_t$, $v_{t}$ a vector of size $3 N_\sat$ defined as 
  $v_{t,i}={\delta_{i,k}}/{\sqrt{N_\sat}}$, and $k \equiv \left[(t-1)\mod3\right]$.

  The phonon frequencies are computed at $\Gamma$ from the corrected Hessian $H_{\rm c}$.
  Following Ref.~\onlinecite{doi:10.1021/acs.chemmater.7b00789}, 
the zero frequencies are replaced by $\left<\omega_\sopt\right>/2$  ($\omega_\sopt$ represent the optical frequencies) in the calculation of vibrational properties. 
  From here, phonon spectral distribution (mean, max, and variance) and thermodynamic properties are then obtained.

\begin{figure}
  \includegraphics[width=0.99\linewidth]{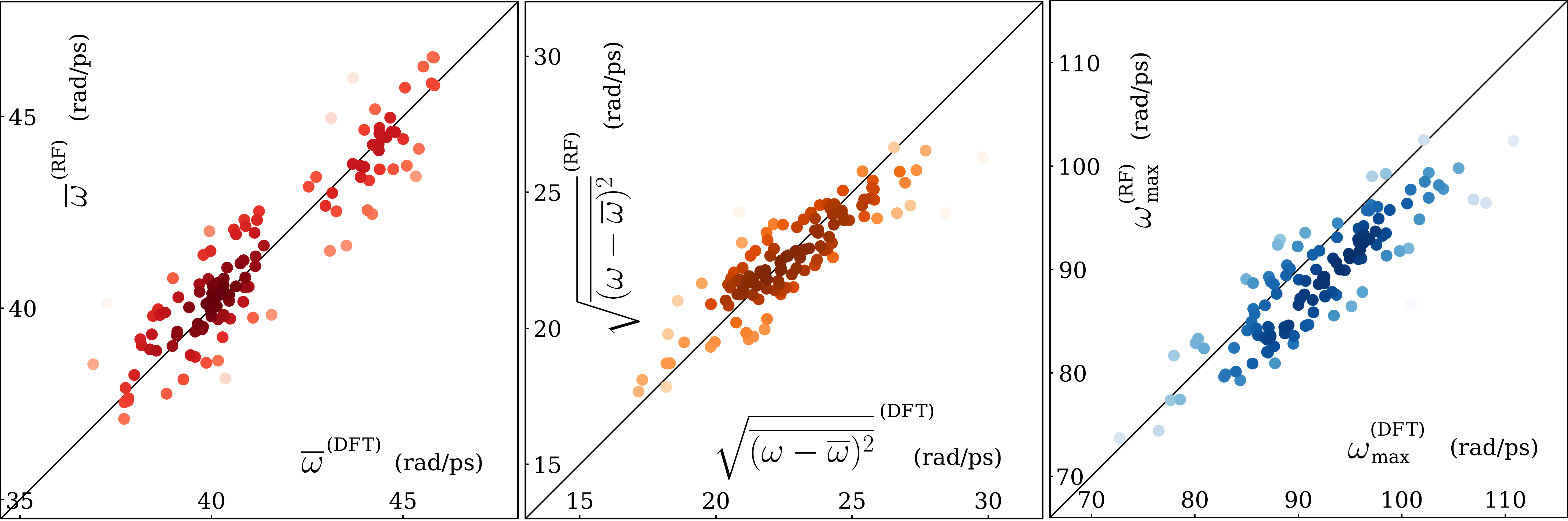}
  \caption{\small Plots of the square root of the variance (middle), the mean (left), and the maximum (right) of the compound frequencies. 
    The $x$-axis shows the values computed with \DFT\ and the $y$-axis shows the values obtained with \RFs.}
  \label{figure4}
\end{figure}

  Figure~\ref{figure4} displays the square root of the variance, the mean, and the maximum of the compound frequencies, computed with \DFT\ and predicted with \RFs. 
  There is good correlation and the statistical analysis is summarized in Table \ref{table1}.

  The specific heats at constant volume, vibrational entropies and free energies are computed at 300 K from the phonon frequencies $\omega$ following Ref. \onlinecite{Landau}
  ($N_{\mathrm{atoms}}$ is the number of atoms in the cell and $n_\omega$ is the Bose-Einstein distribution):
  \begin{eqnarray}
    C_\sv&=&\frac{1}{N_{\mathrm{atoms}}}\sum_{\omega}\frac{\hbar^2\omega^2}{k_{\mathrm{B}}T^2}\frac{\exp\left(\frac{-\hbar\omega}{k_{\mathrm{B}}T}\right)}{\left(1-\exp\left(\frac{-\hbar\omega}{k_{\mathrm{B}}T}\right)\right)^2} \nonumber  \\
    S_\svib&=&\frac{1}{N_{\mathrm{atoms}}}\sum_{\omega}\left\{\frac{\frac{\hbar\omega}{T}\exp\left(\frac{-\hbar\omega}{k_{\mathrm{B}}T}\right)}{1-\exp\left\{\frac{-\hbar\omega}{k_{\mathrm{B}}T}\right)}-k_{\mathrm{B}}\ln{\left[1-\exp\left(\frac{-\hbar\omega}{k_{\mathrm{B}}T}\right)\right]}\right\} \nonumber \\
    F_\svib&=&E_\svib-TS_\svib =-\frac{1}{N_{\mathrm{atoms}}}\sum_{\omega}\left(\frac{\hbar\omega}{2}+k_{\mathrm{B}}T\ln{n_\omega}\right) \nonumber
  \end{eqnarray}

\begin{figure}
  \includegraphics[width=0.99\linewidth]{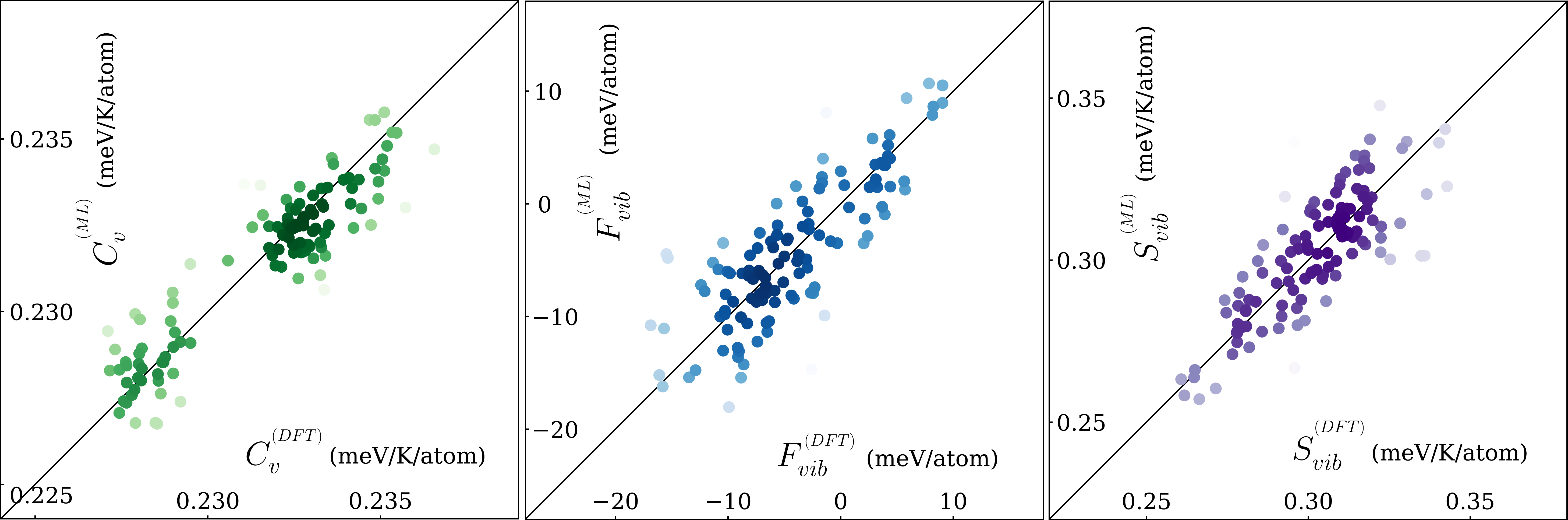}
  \caption{\small Heat capacities ($C_{\rm v}$), vibrational free energies ($F_{\rm vib}$), and vibrational entropies ($S_{\rm vib}$) are computed at 300 K with \DFT\ and \RFs.}
  \label{figure6}
\end{figure}

  The quantities are depicted in Figure~\ref{figure6}: values computed with \DFT\ are on the $x$-axis while values predicted with \RFs\ are on the $y$-axis. 
  The plots show that the \RF\ approach gives a good approximation of the heat capacities, vibrational free energies and vibrational entropies of the different structures of KZnF$_3$. 
  The statistical analysis is summarized in Table \ref{table1}.

  The results  illustrate that descriptors transforming as two-index tensors, in combination with \RFs\ regression algorithms and relatively small training sets, can be used to predict \IFCs\ for generic metastable crystal structures. 
  Other approaches can be used to further improve the outcome:
  amongst them there are 
  support vector machines or neural networks as algorithms,
  different descriptor definitions, inhomogeneous or adjustable grids for the radii,
  replacing the Gaussian by different functions, considering the species as separate descriptors from the structural ones.
  In addition,  training on more and larger supercells, could improve the accuracy and enhance the relevance of 3-atom environmental descriptors, which could also take a different functional form from the presented one. 
  However, regardless of potential improvements, here it has been shown that {\bf it is possible to predict force constants by training exclusively on metastable structures --- very abundant in online respositories --- without the necessity to include unstable configurations}.
  Atomic configurations can be viewed as points in a high-dimensional space of rototranslation-permutation invariant descriptors: close points $\rightarrow$ similar properties. 
  As such, to overcome similarity \ML\ force fields are typically trained on tens of thousands of atomic configurations not corresponding to local energy minima, in order to have a sufficiently dense coverage of the representative hyper-volume in the configuration space:
  any new configuration is close to some other points in the training set, allowing for a good prediction of the energy/forces. 
  In contrast, our case deals only with atomic configurations corresponding to local energy minima: metastable structures are intrinsically dissimilar, belonging to attraction basins separated by energy barriers in the configurational space.
  {\it A priori} there is no reason why the properties of such different structures should be related to each other. 
  {\it A posteriori}, the results show that the Hessians of different local minima are indeed inter-related and strongly determined by pair-wise interactions. 
  This unveils an underlying regularity in the character of the inter-atomic interactions that persists across the different metastable structures, enabling the prediction of force constants of an unknown metastable 
  structure by training only on the other metastable systems available, without having to include any unstable structures to populate the empty configurational space between energy minima. 
  The property can be leveraged for the quick estimation of vibrational contributions to phase stability and transport properties of materials, and to enable the high-throughput {\it ab initio} screening of these properties at finite temperatures.

\section{Conclusions}

  We have shown that Hessians and associated vibrational properties of multi-component metastable structures 
  can be efficiently predicted by machine learning regressions without the need of developing full force fields. 
  The key factors determining the interatomic force constants are captured by tensor descriptors depending only on the species and distance between atoms' pairs.
  The main features of the vibrational spectrum --- maximum, mean and variance --- are correctly reproduced.
  \ML\ predictions of thermodynamic properties --- specific heat, vibrational free energy and entropy --- correlate well with the \DFT\ calculations. 
  Once trained, the model allows for
  the rapid vibrational characterization of relaxed structures with arbitrary complexity at low computational cost
  and the efficient comparison of polymorphs competing for stability at finite temperature.
  It is envisioned that machine learning vibrational-approaches will enable
  the use of the abundant online repositories information for efficient 
  high-throughput screening of stability and transport at finite temperature.

\section{Acknowledgements}
The work is supported by M-era.net through the ICETS project (DFG: MA 5487/4-1 and ANR-14-MERA-0003-03) and ANR through the Carnot MAPPE project.
S.C. acknowledges DOD-ONR (N00014-15-1-2863), the Alexander von Humboldt Foundation and the Max Planck Society for financial support

\begin{doublespacing}
\newcommand{\Ozolins}{Ozoli\c{n}\v{s}}
\providecommand{\latin}[1]{#1}
\providecommand*\mcitethebibliography{\thebibliography}
\csname @ifundefined\endcsname{endmcitethebibliography}
  {\let\endmcitethebibliography\endthebibliography}{}

\end{doublespacing}
\pagebreak

\begin{mcitethebibliography}{41}
\providecommand*\natexlab[1]{#1}
\providecommand*\mciteSetBstSublistMode[1]{}
\providecommand*\mciteSetBstMaxWidthForm[2]{}
\providecommand*\mciteBstWouldAddEndPuncttrue
  {\def\EndOfBibitem{\unskip.}}
\providecommand*\mciteBstWouldAddEndPunctfalse
  {\let\EndOfBibitem\relax}
\providecommand*\mciteSetBstMidEndSepPunct[3]{}
\providecommand*\mciteSetBstSublistLabelBeginEnd[3]{}
\providecommand*\EndOfBibitem{}
\mciteSetBstSublistMode{f}
\mciteSetBstMaxWidthForm{subitem}{(\alph{mcitesubitemcount})}
\mciteSetBstSublistLabelBeginEnd
  {\mcitemaxwidthsubitemform\space}
  {\relax}
  {\relax}

\bibitem[Curtarolo \latin{et~al.}(2012)Curtarolo, Setyawan, Hart, Jahn\'{a}tek,
  Chepulskii, Taylor, Wang, Xue, Yang, Levy, Mehl, Stokes, Demchenko, and
  Morgan]{aflowPAPER}
Curtarolo,~S.; Setyawan,~W.; Hart,~G.~L.~W.; Jahn\'{a}tek,~M.;
  Chepulskii,~R.~V.; Taylor,~R.~H.; Wang,~S.; Xue,~J.; Yang,~K.; Levy,~O.;
  Mehl,~M.~J.; Stokes,~H.~T.; Demchenko,~D.~O.; Morgan,~D. {AFLOW}: An
  automatic framework for high-throughput materials discovery. \emph{Comput.\
  Mater.\ Sci.} \textbf{2012}, \emph{58}, 218--226\relax
\mciteBstWouldAddEndPuncttrue
\mciteSetBstMidEndSepPunct{\mcitedefaultmidpunct}
{\mcitedefaultendpunct}{\mcitedefaultseppunct}\relax
\EndOfBibitem
\bibitem[Curtarolo \latin{et~al.}(2012)Curtarolo, Setyawan, Wang, Xue, Yang,
  Taylor, Nelson, Hart, Sanvito, {Buongiorno Nardelli}, Mingo, and
  Levy]{aflowlibPAPER}
Curtarolo,~S.; Setyawan,~W.; Wang,~S.; Xue,~J.; Yang,~K.; Taylor,~R.~H.;
  Nelson,~L.~J.; Hart,~G.~L.~W.; Sanvito,~S.; {Buongiorno Nardelli},~M.;
  Mingo,~N.; Levy,~O. {AFLOWLIB.ORG}: A distributed materials properties
  repository from high-throughput {\it ab initio} calculations. \emph{Comput.\
  Mater.\ Sci.} \textbf{2012}, \emph{58}, 227--235\relax
\mciteBstWouldAddEndPuncttrue
\mciteSetBstMidEndSepPunct{\mcitedefaultmidpunct}
{\mcitedefaultendpunct}{\mcitedefaultseppunct}\relax
\EndOfBibitem
\bibitem[Jain \latin{et~al.}(2013)Jain, Ong, Hautier, Chen, Richards, Dacek,
  Cholia, Gunter, Skinner, Ceder, and Persson]{Jain2013}
Jain,~A.; Ong,~S.~P.; Hautier,~G.; Chen,~W.; Richards,~W.~D.; Dacek,~S.;
  Cholia,~S.; Gunter,~D.; Skinner,~D.; Ceder,~G.; Persson,~K.~a. {The Materials
  Project: A materials genome approach to accelerating materials innovation}.
  \emph{APL Materials} \textbf{2013}, \emph{1}, 011002\relax
\mciteBstWouldAddEndPuncttrue
\mciteSetBstMidEndSepPunct{\mcitedefaultmidpunct}
{\mcitedefaultendpunct}{\mcitedefaultseppunct}\relax
\EndOfBibitem
\bibitem[Kirklin \latin{et~al.}(2015)Kirklin, Saal, Meredig, Thompson, Doak,
  Aykol, Rühl, and Wolverton]{OQMD}
Kirklin,~S.; Saal,~J.~E.; Meredig,~B.; Thompson,~A.; Doak,~J.~W.; Aykol,~M.;
  Rühl,~S.; Wolverton,~C. {The Open Quantum Materials Database (OQMD):
  assessing the accuracy of DFT formation energies}. \emph{Npj Computational
  Materials} \textbf{2015}, \emph{1}, 15010\relax
\mciteBstWouldAddEndPuncttrue
\mciteSetBstMidEndSepPunct{\mcitedefaultmidpunct}
{\mcitedefaultendpunct}{\mcitedefaultseppunct}\relax
\EndOfBibitem
\bibitem[Nosengo(2016)]{nosengo_can_2016}
Nosengo,~N. Can artificial intelligence create the next wonder material?
  \emph{Nature} \textbf{2016}, \emph{533}, 22--25\relax
\mciteBstWouldAddEndPuncttrue
\mciteSetBstMidEndSepPunct{\mcitedefaultmidpunct}
{\mcitedefaultendpunct}{\mcitedefaultseppunct}\relax
\EndOfBibitem
\bibitem[Curtarolo \latin{et~al.}(2013)Curtarolo, Hart, Nardelli, Mingo,
  Sanvito, and Levy]{Thehigh-throughputhighwaytocomputationalmaterialsdesign}
Curtarolo,~S.; Hart,~G. L.~W.; Nardelli,~M.~B.; Mingo,~N.; Sanvito,~S.;
  Levy,~O. The High-Throughput Highway to Computational Materials Design.
  \emph{Nat. Mater.} \textbf{2013}, \emph{12}, 191--201\relax
\mciteBstWouldAddEndPuncttrue
\mciteSetBstMidEndSepPunct{\mcitedefaultmidpunct}
{\mcitedefaultendpunct}{\mcitedefaultseppunct}\relax
\EndOfBibitem
\bibitem[Green \latin{et~al.}(2017)Green, Choi, Hattrick-Simpers, Joshi,
  Takeuchi, Barron, Campo, Chiang, Empedocles, Gregoire, Kusne, Martin, Mehta,
  Persson, Trautt, Duren, and Zakutayev]{doi:10.1063/1.4977487}
Green,~M.~L. \latin{et~al.}  Fulfilling the promise of the materials genome
  initiative with high-throughput experimental methodologies. \emph{Applied
  Physics Reviews} \textbf{2017}, \emph{4}, 011105\relax
\mciteBstWouldAddEndPuncttrue
\mciteSetBstMidEndSepPunct{\mcitedefaultmidpunct}
{\mcitedefaultendpunct}{\mcitedefaultseppunct}\relax
\EndOfBibitem
\bibitem[Toher \latin{et~al.}(2017)Toher, Oses, Plata, Hicks, Rose, Levy, {de
  Jong}, Asta, Fornari, {Buongiorno Nardelli}, and Curtarolo]{curtarolo:art115}
Toher,~C.; Oses,~C.; Plata,~J.~J.; Hicks,~D.; Rose,~F.; Levy,~O.; {de
  Jong},~M.; Asta,~M.~D.; Fornari,~M.; {Buongiorno Nardelli},~M.; Curtarolo,~S.
  Combining the {AFLOW} {GIBBS} and Elastic Libraries to efficiently and
  robustly screen thermomechanical properties of solids. \emph{Phys.\ Rev.\
  Mater.} \textbf{2017}, \emph{1}, 015401\relax
\mciteBstWouldAddEndPuncttrue
\mciteSetBstMidEndSepPunct{\mcitedefaultmidpunct}
{\mcitedefaultendpunct}{\mcitedefaultseppunct}\relax
\EndOfBibitem
\bibitem[Opahle \latin{et~al.}(2013)Opahle, Parma, McEniry, Drautz, and
  Madsen]{Opahle_NJP13}
Opahle,~I.; Parma,~A.; McEniry,~E.~J.; Drautz,~R.; Madsen,~G. K.~H.
  High-throughput study of the structural stability and thermoelectric
  properties of transition metal silicides. \emph{New Journal of Physics}
  \textbf{2013}, \emph{15}, 105010\relax
\mciteBstWouldAddEndPuncttrue
\mciteSetBstMidEndSepPunct{\mcitedefaultmidpunct}
{\mcitedefaultendpunct}{\mcitedefaultseppunct}\relax
\EndOfBibitem
\bibitem[Sun \latin{et~al.}(2016)Sun, Dacek, Ong, Hautier, Jain, Richards,
  Gamst, Persson, and Ceder]{sun_thermodynamic_2016}
Sun,~W.; Dacek,~S.~T.; Ong,~S.~P.; Hautier,~G.; Jain,~A.; Richards,~W.~D.;
  Gamst,~A.~C.; Persson,~K.~A.; Ceder,~G. The thermodynamic scale of inorganic
  crystalline metastability. \emph{Science Advances} \textbf{2016},
  \emph{2}\relax
\mciteBstWouldAddEndPuncttrue
\mciteSetBstMidEndSepPunct{\mcitedefaultmidpunct}
{\mcitedefaultendpunct}{\mcitedefaultseppunct}\relax
\EndOfBibitem
\bibitem[Sun \latin{et~al.}(2016)Sun, Remsing, Zhang, Sun, Ruzsinszky, Peng,
  Yang, Paul, Waghmare, Wu, Klein, and Perdew]{Sun_NatChem16}
Sun,~J.; Remsing,~R.~C.; Zhang,~Y.; Sun,~Z.; Ruzsinszky,~A.; Peng,~H.;
  Yang,~Z.; Paul,~A.; Waghmare,~U.; Wu,~X.; Klein,~M.~L.; Perdew,~J.~P.
  {Accurate first-principles structures and energies of diversely bonded
  systems from an efficient density functional}. \emph{Nature Chem.}
  \textbf{2016}, \emph{{8}}, {831--836}\relax
\mciteBstWouldAddEndPuncttrue
\mciteSetBstMidEndSepPunct{\mcitedefaultmidpunct}
{\mcitedefaultendpunct}{\mcitedefaultseppunct}\relax
\EndOfBibitem
\bibitem[Liu \latin{et~al.}(2017)Liu, Burton, Khare, and
  Gall]{liu_first-principles_2017}
Liu,~Z. T.~Y.; Burton,~B.~P.; Khare,~S.~V.; Gall,~D. First-principles phase
  diagram calculations for the rocksalt-structure quasibinary systems
  {TiN}â€“{ZrN}, {TiN}â€“{HfN} and {ZrN}â€“{HfN}. \emph{Journal
  of Physics: Condensed Matter} \textbf{2017}, \emph{29}, 035401\relax
\mciteBstWouldAddEndPuncttrue
\mciteSetBstMidEndSepPunct{\mcitedefaultmidpunct}
{\mcitedefaultendpunct}{\mcitedefaultseppunct}\relax
\EndOfBibitem
\bibitem[Burton and van~de Walle(2006)Burton, and van~de
  Walle]{burton_first-principles_2006}
Burton,~B.~P.; van~de Walle,~A. First-principles phase diagram calculations for
  the system {NaCl}â€“{KCl}: {The} role of excess vibrational entropy.
  \emph{Chemical Geology} \textbf{2006}, \emph{225}, 222--229\relax
\mciteBstWouldAddEndPuncttrue
\mciteSetBstMidEndSepPunct{\mcitedefaultmidpunct}
{\mcitedefaultendpunct}{\mcitedefaultseppunct}\relax
\EndOfBibitem
\bibitem[R.~Akbarzadeh \latin{et~al.}(2007)R.~Akbarzadeh, OzoliÅ†Å¡, and
  Wolverton]{r._akbarzadeh_first-principles_2007}
R.~Akbarzadeh,~A.; OzoliÅ†Å¡,~V.; Wolverton,~C. First-{Principles}
  {Determination} of {Multicomponent} {Hydride} {Phase} {Diagrams}:
  {Application} to the {Li}-{Mg}-{N}-{H} {System}. \emph{Advanced Materials}
  \textbf{2007}, \emph{19}, 3233--3239\relax
\mciteBstWouldAddEndPuncttrue
\mciteSetBstMidEndSepPunct{\mcitedefaultmidpunct}
{\mcitedefaultendpunct}{\mcitedefaultseppunct}\relax
\EndOfBibitem
\bibitem[Carrete \latin{et~al.}(2017)Carrete, Gallego, and
  Mingo]{carrete_structural_2017}
Carrete,~J.; Gallego,~L.~J.; Mingo,~N. Structural {Complexity} and {Phonon}
  {Physics} in 2D {Arsenenes}. \emph{The Journal of Physical Chemistry Letters}
  \textbf{2017}, 1375--1380\relax
\mciteBstWouldAddEndPuncttrue
\mciteSetBstMidEndSepPunct{\mcitedefaultmidpunct}
{\mcitedefaultendpunct}{\mcitedefaultseppunct}\relax
\EndOfBibitem
\bibitem[Körmann \latin{et~al.}(2017)Körmann, Ikeda, Grabowski, and
  Sluiter]{phononbroadeninginHEA}
Körmann,~F.; Ikeda,~Y.; Grabowski,~B.; Sluiter,~M. H.~F. Phonon broadening in
  high entropy alloys. \emph{npj Computational Materials} \textbf{2017},
  \emph{3}, 36\relax
\mciteBstWouldAddEndPuncttrue
\mciteSetBstMidEndSepPunct{\mcitedefaultmidpunct}
{\mcitedefaultendpunct}{\mcitedefaultseppunct}\relax
\EndOfBibitem
\bibitem[Oganov and Glass(2006)Oganov, and Glass]{doi:10.1063/1.2210932}
Oganov,~A.~R.; Glass,~C.~W. Crystal structure prediction using ab initio
  evolutionary techniques: Principles and applications. \emph{The Journal of
  Chemical Physics} \textbf{2006}, \emph{124}, 244704\relax
\mciteBstWouldAddEndPuncttrue
\mciteSetBstMidEndSepPunct{\mcitedefaultmidpunct}
{\mcitedefaultendpunct}{\mcitedefaultseppunct}\relax
\EndOfBibitem
\bibitem[Glass \latin{et~al.}(2006)Glass, Oganov, and Hansen]{GLASS2006713}
Glass,~C.~W.; Oganov,~A.~R.; Hansen,~N. USPEX — Evolutionary crystal
  structure prediction. \emph{Computer Physics Communications} \textbf{2006},
  \emph{175}, 713 -- 720\relax
\mciteBstWouldAddEndPuncttrue
\mciteSetBstMidEndSepPunct{\mcitedefaultmidpunct}
{\mcitedefaultendpunct}{\mcitedefaultseppunct}\relax
\EndOfBibitem
\bibitem[Wolverton and \Ozolins(2001)Wolverton, and
  \Ozolins]{wolverton:prl_2001_AlCu}
Wolverton,~C.; \Ozolins,~V. Entropically Favored Ordering: The Metallurgy of
  {Al}$_2${Cu} Revisited. \emph{Phys.\ Rev.\ Lett.} \textbf{2001}, \emph{86},
  5518\relax
\mciteBstWouldAddEndPuncttrue
\mciteSetBstMidEndSepPunct{\mcitedefaultmidpunct}
{\mcitedefaultendpunct}{\mcitedefaultseppunct}\relax
\EndOfBibitem
\bibitem[Nyman and Day(2015)Nyman, and Day]{C5CE00045A}
Nyman,~J.; Day,~G.~M. Static and lattice vibrational energy differences between
  polymorphs. \emph{CrystEngComm} \textbf{2015}, \emph{17}, 5154--5165\relax
\mciteBstWouldAddEndPuncttrue
\mciteSetBstMidEndSepPunct{\mcitedefaultmidpunct}
{\mcitedefaultendpunct}{\mcitedefaultseppunct}\relax
\EndOfBibitem
\bibitem[Bergerhoff \latin{et~al.}(1983)Bergerhoff, Hundt, Sievers, and
  Brown]{ICSD}
Bergerhoff,~G.; Hundt,~R.; Sievers,~R.; Brown,~I.~D. The inorganic crystal
  structure data base. \emph{J.\ Chem.\ Inf.\ Comput.\ Sci.} \textbf{1983},
  \emph{23}, 66--69\relax
\mciteBstWouldAddEndPuncttrue
\mciteSetBstMidEndSepPunct{\mcitedefaultmidpunct}
{\mcitedefaultendpunct}{\mcitedefaultseppunct}\relax
\EndOfBibitem
\bibitem[van~de Walle and Ceder(2002)van~de Walle, and
  Ceder]{van_de_walle_effect_2002}
van~de Walle,~A.; Ceder,~G. The effect of lattice vibrations on substitutional
  alloy thermodynamics. \emph{Reviews of Modern Physics} \textbf{2002},
  \emph{74}, 11--45\relax
\mciteBstWouldAddEndPuncttrue
\mciteSetBstMidEndSepPunct{\mcitedefaultmidpunct}
{\mcitedefaultendpunct}{\mcitedefaultseppunct}\relax
\EndOfBibitem
\bibitem[Curtarolo \latin{et~al.}(2005)Curtarolo, Morgan, and
  Ceder]{curtarolo_accuracy_2005}
Curtarolo,~S.; Morgan,~D.; Ceder,~G. Accuracy of ab initio methods in
  predicting the crystal structures of metals: {A} review of 80 binary alloys.
  \emph{Calphad} \textbf{2005}, \emph{29}, 163--211\relax
\mciteBstWouldAddEndPuncttrue
\mciteSetBstMidEndSepPunct{\mcitedefaultmidpunct}
{\mcitedefaultendpunct}{\mcitedefaultseppunct}\relax
\EndOfBibitem
\bibitem[Carrete \latin{et~al.}(2014)Carrete, Li, Mingo, Wang, and
  Curtarolo]{PhysRevX.4.011019}
Carrete,~J.; Li,~W.; Mingo,~N.; Wang,~S.; Curtarolo,~S. Finding Unprecedentedly
  Low-Thermal-Conductivity Half-Heusler Semiconductors via High-Throughput
  Materials Modeling. \emph{Phys. Rev. X} \textbf{2014}, \emph{4}, 011019\relax
\mciteBstWouldAddEndPuncttrue
\mciteSetBstMidEndSepPunct{\mcitedefaultmidpunct}
{\mcitedefaultendpunct}{\mcitedefaultseppunct}\relax
\EndOfBibitem
\bibitem[van Roekeghem \latin{et~al.}(2016)van Roekeghem, Carrete, Oses,
  Curtarolo, and Mingo]{PhysRevX.6.041061}
van Roekeghem,~A.; Carrete,~J.; Oses,~C.; Curtarolo,~S.; Mingo,~N.
  High-Throughput Computation of Thermal Conductivity of High-Temperature Solid
  Phases: The Case of Oxide and Fluoride Perovskites. \emph{Phys. Rev. X}
  \textbf{2016}, \emph{6}, 041061\relax
\mciteBstWouldAddEndPuncttrue
\mciteSetBstMidEndSepPunct{\mcitedefaultmidpunct}
{\mcitedefaultendpunct}{\mcitedefaultseppunct}\relax
\EndOfBibitem
\bibitem[Legrain \latin{et~al.}(2017)Legrain, Carrete, van Roekeghem,
  Curtarolo, and Mingo]{doi:10.1021/acs.chemmater.7b00789}
Legrain,~F.; Carrete,~J.; van Roekeghem,~A.; Curtarolo,~S.; Mingo,~N. How
  Chemical Composition Alone Can Predict Vibrational Free Energies and
  Entropies of Solids. \emph{Chemistry of Materials} \textbf{2017}, \emph{29},
  6220--6227\relax
\mciteBstWouldAddEndPuncttrue
\mciteSetBstMidEndSepPunct{\mcitedefaultmidpunct}
{\mcitedefaultendpunct}{\mcitedefaultseppunct}\relax
\EndOfBibitem
\bibitem[Behler(2011)]{doi:10.1063/1.3553717}
Behler,~J. Atom-centered symmetry functions for constructing high-dimensional
  neural network potentials. \emph{The Journal of Chemical Physics}
  \textbf{2011}, \emph{134}, 074106\relax
\mciteBstWouldAddEndPuncttrue
\mciteSetBstMidEndSepPunct{\mcitedefaultmidpunct}
{\mcitedefaultendpunct}{\mcitedefaultseppunct}\relax
\EndOfBibitem
\bibitem[Botu and Ramprasad(2015)Botu, and Ramprasad]{PhysRevB.92.094306}
Botu,~V.; Ramprasad,~R. Learning scheme to predict atomic forces and accelerate
  materials simulations. \emph{Phys. Rev. B} \textbf{2015}, \emph{92},
  094306\relax
\mciteBstWouldAddEndPuncttrue
\mciteSetBstMidEndSepPunct{\mcitedefaultmidpunct}
{\mcitedefaultendpunct}{\mcitedefaultseppunct}\relax
\EndOfBibitem
\bibitem[Botu and Ramprasad(2015)Botu, and Ramprasad]{QUA:QUA24836}
Botu,~V.; Ramprasad,~R. Adaptive machine learning framework to accelerate ab
  initio molecular dynamics. \emph{International Journal of Quantum Chemistry}
  \textbf{2015}, \emph{115}, 1074--1083\relax
\mciteBstWouldAddEndPuncttrue
\mciteSetBstMidEndSepPunct{\mcitedefaultmidpunct}
{\mcitedefaultendpunct}{\mcitedefaultseppunct}\relax
\EndOfBibitem
\bibitem[Artrith \latin{et~al.}(2017)Artrith, Urban, and
  Ceder]{artrith_efficient_2017}
Artrith,~N.; Urban,~A.; Ceder,~G. Efficient and accurate machine-learning
  interpolation of atomic energies in compositions with many species.
  \emph{Physical Review B} \textbf{2017}, \emph{96}, 014112\relax
\mciteBstWouldAddEndPuncttrue
\mciteSetBstMidEndSepPunct{\mcitedefaultmidpunct}
{\mcitedefaultendpunct}{\mcitedefaultseppunct}\relax
\EndOfBibitem
\bibitem[Hohenberg and Kohn(1964)Hohenberg, and Kohn]{InhomogeneousElectronGas}
Hohenberg,~P.; Kohn,~W. Inhomogeneous Electron Gas. \emph{Phys. Rev.}
  \textbf{1964}, \emph{136}, B864--B871\relax
\mciteBstWouldAddEndPuncttrue
\mciteSetBstMidEndSepPunct{\mcitedefaultmidpunct}
{\mcitedefaultendpunct}{\mcitedefaultseppunct}\relax
\EndOfBibitem
\bibitem[Kohn and Sham(1965)Kohn, and Sham]{PhysRev.140.A1133}
Kohn,~W.; Sham,~L.~J. Self-Consistent Equations Including Exchange and
  Correlation Effects. \emph{Phys. Rev.} \textbf{1965}, \emph{140},
  A1133--A1138\relax
\mciteBstWouldAddEndPuncttrue
\mciteSetBstMidEndSepPunct{\mcitedefaultmidpunct}
{\mcitedefaultendpunct}{\mcitedefaultseppunct}\relax
\EndOfBibitem
\bibitem[Kresse and Furthm\"{u}ller(1996)Kresse, and
  Furthm\"{u}ller]{Efficiencyofab-initiototalenergy}
Kresse,~G.; Furthm\"{u}ller,~J. Efficiency of Ab-Initio Total Energy
  Calculations for Metals and Semiconductors Using a Plane-Wave Basis Set.
  \emph{Comput. Mater. Sci.} \textbf{1996}, \emph{6}, 15 -- 50\relax
\mciteBstWouldAddEndPuncttrue
\mciteSetBstMidEndSepPunct{\mcitedefaultmidpunct}
{\mcitedefaultendpunct}{\mcitedefaultseppunct}\relax
\EndOfBibitem
\bibitem[Kresse and Joubert(1999)Kresse, and
  Joubert]{Fromultrasoftpseudopotentials}
Kresse,~G.; Joubert,~D. From Ultrasoft Pseudopotentials to the Projector
  Augmented-Wave Method. \emph{Phys. Rev. B} \textbf{1999}, \emph{59},
  1758--1775\relax
\mciteBstWouldAddEndPuncttrue
\mciteSetBstMidEndSepPunct{\mcitedefaultmidpunct}
{\mcitedefaultendpunct}{\mcitedefaultseppunct}\relax
\EndOfBibitem
\bibitem[Taylor \latin{et~al.}(2014)Taylor, Rose, Toher, Levy, Yang,
  {Buongiorno Nardelli}, and Curtarolo]{RESTful_API_2014}
Taylor,~R.~H.; Rose,~F.; Toher,~C.; Levy,~O.; Yang,~K.; {Buongiorno
  Nardelli},~M.; Curtarolo,~S. A {REST}ful {API} for exchanging materials data
  in the {AFLOWLIB}.org consortium. \emph{Comput.\ Mater.\ Sci.} \textbf{2014},
  \emph{93}, 178--192\relax
\mciteBstWouldAddEndPuncttrue
\mciteSetBstMidEndSepPunct{\mcitedefaultmidpunct}
{\mcitedefaultendpunct}{\mcitedefaultseppunct}\relax
\EndOfBibitem
\bibitem[Calderon \latin{et~al.}(2015)Calderon, Plata, Toher, Oses, Levy,
  Fornari, Natan, Mehl, Hart, {Buongiorno Nardelli}, and
  Curtarolo]{curtarolo:art104}
Calderon,~C.~E.; Plata,~J.~J.; Toher,~C.; Oses,~C.; Levy,~O.; Fornari,~M.;
  Natan,~A.; Mehl,~M.~J.; Hart,~G.~L.~W.; {Buongiorno Nardelli},~M.;
  Curtarolo,~S. The {AFLOW} standard for high-throughput materials science
  calculations. \emph{Comput.\ Mater.\ Sci.} \textbf{2015}, \emph{108 Part A},
  233--238\relax
\mciteBstWouldAddEndPuncttrue
\mciteSetBstMidEndSepPunct{\mcitedefaultmidpunct}
{\mcitedefaultendpunct}{\mcitedefaultseppunct}\relax
\EndOfBibitem
\bibitem[Baroni \latin{et~al.}(2001)Baroni, de~Gironcoli, Dal~Corso, and
  Giannozzi]{DFTP}
Baroni,~S.; de~Gironcoli,~S.; Dal~Corso,~A.; Giannozzi,~P. Phonons and Related
  Crystal Properties from Density-Functional Perturbation Theory. \emph{Rev.
  Mod. Phys.} \textbf{2001}, \emph{73}, 515--562\relax
\mciteBstWouldAddEndPuncttrue
\mciteSetBstMidEndSepPunct{\mcitedefaultmidpunct}
{\mcitedefaultendpunct}{\mcitedefaultseppunct}\relax
\EndOfBibitem
\bibitem[Pedregosa \latin{et~al.}(2011)Pedregosa, Varoquaux, Gramfort, Michel,
  Thirion, Grisel, Blondel, Prettenhofer, Weiss, Dubourg, Vanderplas, Passos,
  Cournapeau, Brucher, Perrot, and Duchesnay]{Pedregosa_JMLR_2011}
Pedregosa,~F. \latin{et~al.}  Scikit-learn: Machine Learning in {P}ython.
  \emph{J.\ Mach.\ Learn.\ Res.} \textbf{2011}, \relax
\mciteBstWouldAddEndPunctfalse
\mciteSetBstMidEndSepPunct{\mcitedefaultmidpunct}
{}{\mcitedefaultseppunct}\relax
\EndOfBibitem
\bibitem[Spencer(1980)]{Longman}
Spencer,~A. J.~M. \emph{Continuum Mechanics}; Longman Scientific and Technical,
  1980\relax
\mciteBstWouldAddEndPuncttrue
\mciteSetBstMidEndSepPunct{\mcitedefaultmidpunct}
{\mcitedefaultendpunct}{\mcitedefaultseppunct}\relax
\EndOfBibitem
\bibitem[Landau and Lifshitz(1969)Landau, and Lifshitz]{Landau}
Landau,~L.~D.; Lifshitz,~E.~M. \emph{Statistical Physics (Second Revised and
  Enlarged Edition)}; Pergamon Press: Oxford, 1969\relax
\mciteBstWouldAddEndPuncttrue
\mciteSetBstMidEndSepPunct{\mcitedefaultmidpunct}
{\mcitedefaultendpunct}{\mcitedefaultseppunct}\relax
\EndOfBibitem
\end{mcitethebibliography}
\end{document}